# Identifying substitutional oxygen as a prolific point defect in monolayer transition metal dichalcogenides with experiment and theory

Sara Barja[1,2,3,4^*], Sivan Refaely-Abramson[1,5^], Bruno Schuler[1^], Diana Y. Qiu[5,6^], Artem Pulkin[7], Sebastian Wickenburg[1], Hyejin Ryu[8,9], Miguel M. Ugeda[2,3,4], Christoph Kastl[1], Christopher Chen[1], Choongyu Hwang[10], Adam Schwartzberg[1], Shaul Aloni[1], Sung-Kwan Mo[8], D. Frank Ogletree[1], Michael F. Crommie[5,11], Oleg V. Yazyev[7], Steven G. Louie[5,6*], Jeffrey B. Neaton[1,5,11]* and Alexander Weber-Bargioni[1]*

[1] Molecular Foundry, Lawrence Berkeley National Laboratory, California 94720, USA.
[2] Departamento de Física de Materiales, Centro de Física de Materiales, University of the Basque Country UPV/EHU-CSIC, Donostia-San Sebastián 20018, Spain
[3] IKERBASQUE, Basque Foundation for Science
[4] Donostia International Physics Center, Donostia-San Sebastián 20018, Spain
[5] Department of Physics, University of California at Berkeley, Berkeley, California 94720, USA.
[6] Materials Sciences Division, Lawrence Berkeley National Laboratory, California 94720, USA
[7] Institute of Physics, Ecole Polytechnique Fédérale de Lausanne (EPFL), CH-1015 Lausanne, Switzerland
[8] Advanced Light Source, Lawrence Berkeley National Laboratory, Berkeley, CA 94720, USA.
[9] Center for Spintronics, Korea Institute of Science and Technology, Seoul 02792, Korea
[10] Department of Physics, Pusan National University, Busan 46241, Korea
[11] Kavli Energy NanoSciences Institute at the University of California Berkeley and the Lawrence Berkeley National Laboratory, Berkeley, California 94720, USA.
^Equally contributed
* Corresponding author: sara.barja@ehu.eus, sglouie@berkeley.edu, jbneaton@lbl.gov and afweber-bargioni@lbl.gov.

**Abstract**

Chalcogen vacancies are generally considered to be the most common point defects in transition metal dichalcogenide (TMD) semiconductors because of their low formation energy in vacuum and their frequent observation in transmission electron microscopy studies. Consequently, unexpected optical, transport, and catalytic properties observed in 2D-TMDs have been attributed to deep in-gap states associated with point vacancies, even in the absence of direct experimental evidence. Here, we combine low-temperature non-contact atomic force microscopy, scanning tunneling microscopy and spectroscopy, and state-of-the-art ab initio density functional theory and GW calculations to determine both the atomic structure and electronic properties of the most abundant chalcogen-site point defect common to 2D-TMD semiconductors. Surprisingly, we observe no in-gap states for the chalcogen defects. Our results and analysis strongly suggest that the common chalcogen defects in monolayer $MoSe_2$ and monolayer $WS_2$, prepared and measured in standard environments, are substitutional defects, where a chalcogen atom is substituted by an oxygen atom, rather than vacancies.

**Introduction**

Crystal defects are known to modify semiconductor functionality and are expected to have particularly strong impact on the properties of two-dimensional (2D) materials, where screening is reduced compared to bulk systems[1]. In particular, 2D transition metal dichalcogenides (TMDs) can feature a variety of different defect geometries and related electronic states[2,3]. Consequently, correlating individual structural defects with electronic properties is key for understanding the behavior of and, ultimately, engineering of functional 2D-TMDs. However, the experimental identification of individual defects and the direct correlation of these measurements to their electronic structure is challenging.

Chalcogen vacancies are believed to be the most abundant point defects in 2D-TMD semiconductors, and they are theoretically predicted to introduce deep in-gap states (IGS)[4–10]. As a result, important features in the experimental transport characteristic[9], optical response[5,8,10–13] and catalytic activity[14–16] of 2D-TMDs have typically been attributed to chalcogen vacancies, based on indirect support from images acquired by transmission electron microscopy (TEM)[6,9,11,14,17,18] and scanning tunneling microscopy (STM)[13,16,19–21]. However, TEM measurements does not provide direct access to the electronic structure of individual defects. The difficulties in discriminating from the native point defects and those created by TEM due to radiation damage effects has been widely reported[3,22]. Furthermore, light substitutional atoms, such as oxygen, will produce only very weak TEM contrast and could be mistaken for vacancies[23]. Altogether, these challenges limit the direct correlation of TEM studies on TMD materials with macroscopic response, and the optimization of the material's performance if based on these results. Non-invasive scanning tunneling microscopy (STM) has been used to study both the structure and the electronic properties of point defects in 2D-TMDs. While scanning tunneling spectroscopy (STS) is a direct probe of local electronic structure the individual defects, the interpretation of their chemical nature of atomically-resolved STM images in 2D-TMDs is ambiguous[24] due to the convolution of geometric and electronic structure, which is particularly complex for semiconductors. Prominent features in previous STM images are commonly attributed to chalcogen atom positions[24,25]. Presumably, apparent depressions in the STM images vacancies have been as assigned to chalcogen vacancies in 2D-$MoS_2$[16,19–21] and 2D-$TiSe_2$[26,27], while reported as W vacancies in $WSe_2$ samples[13], guided by the absence of IGS in STS measurements. The strong dependence of the tunneling conditions on the STM contrast of the atomic lattice and the unsatisfactory differentiation between chalcogen and metal sublattices in former STM studies has led to a non-consistent interpretation of the defect type across the current literature[13,14,16,19–21,26,27].

Sulphur vacancies, and their corresponding IGS, have been held responsible for unexpected catalytic activity in hydrogen evolution reactions reported in $MoS_2$[14,15]. This assumption is challenged, however, by the enhanced catalytic activity reported after long-term ambient exposure of $MoS_2$ samples[16], since this enhanced catalysis is attributed to oxygen substitution of S atoms, which have been predicted to lack any IGS[18,28,29]. In order to achieve a fundamental understanding of the effect of defects on the electronic structure, a direct correlation between the atomic and electronic structure of individual defects in 2D-TMDs is required.

Combining complementary techniques that allow access to the material at the atomic-scale – low-temperature non-contact atomic force microscopy (nc-AFM), STM and STS - together with parallel state-of-the-art first-principles ground- and excited-state calculations - using DFT and many-body perturbation theory within the GW approach, respectively – enables a comprehensive interrogation of the system. In this work, we demonstrate how the combination of these methods can reveal the structure of the most abundant type of defect in our 2D-$MoSe_2$ and 2D-$WS_2$ samples. We directly relate atomic and electronic structure through combined nc-AFM and STM/STS measurements of individual point defects in monolayer $MoSe_2$ grown by molecular beam epitaxy (MBE) and in monolayer $WS_2$ grown by chemical vapor deposition (CVD) (see Methods). Although our nc-AFM and STM images of chalcogen defects appear to be consistent with vacancies, a comparison with our DFT and GW calculations can establish these defects as substitutional oxygen, consistent with the lack of IGS. Our comprehensive joint experimental and theoretical study reveals substitutional oxygen as a prolific point defect in 2D-TMDs and provides critical insight for future defect engineering in these systems.

## Results

**Structural characterization of point defects in 2D-MoSe$_2$ and 2D-WS$_2$.**

Large-scale STM images measured on single layer of MoSe$_2$ and WS$_2$ show predominantly two types of point defects structures (see Supplementary Information, Figure S1 a). Figure 1a shows a nc-AFM image of the most abundant types of point defects imaged in our 2D-MoSe$_2$ samples, measured using a CO-functionalized tip for enhanced spatial resolution[30]. Based on the known contrast mechanism of CO-tip nc-AFM[30–32] we assign the hexagonal lattice of bright features (higher frequency shifts) to the outer chalcogen atoms, which are close enough to the tip to generate repulsive forces, and the dark features (lower frequency shifts) to the lower-lying metal atoms, whose larger distance from the tip resulted in purely attractive forces. This assignment unambiguously identifies the lattice sites of MoSe$_2$[33], outlined in Fig. 1a. Accordingly, the two main defect features we observe are located on Se-sublattice sites. Figures 1b and c show the STM constant current images of the exact same pair of defects than in Fig. 1a in the valence band (VB; $V_{sample}$= -1.55V) and conduction band (CB; $V_{sample}$= +0.7V), respectively. The defect-induced modification of the local density of states (LDOS) for both the VB and CB shows a distinct three-fold symmetry with a spatial extent of about 2 nm. States associated with the corresponding defects in 2D-WS$_2$ measured at energies around the CB onset exhibit an apparent six-fold symmetry (see Supplementary Information, Figure S1). States with six-fold symmetry are also observed for the equivalent defects in 2D-MoSe$_2$ at a sample bias of about 200 mV above the onset of the CB energy (see Supplementary Information, Figure S2 b and d).

**Electronic characterization of point defects in 2D-MoSe$_2$ and 2D-WS$_2$.**

Structural defects are well-known to alter the electronic structure of a semiconductor via additional localized states. We study the LDOS of the same type of point defects in MoSe$_2$ and WS$_2$ using STM dI/dV spectroscopy. Figure 2a shows a representative dI/dV spectrum (red line) measured on top of the left point defect in Fig. 1. The spectrum reveals a well-defined bandgap - equivalent to the gap measured on pristine 2D-MoSe$_2$[24,33] at a location next to the defect (black line) - and lacks IGS. An additional defect state is visible about 300 mV below the VB edge. dI/dV spectra measured on the right point defect in Fig.1 also lacks IGS and exhibits the same qualitative spectroscopic features as the dI/dV spectra measured on the left defect (see Supplementary Information, Figure S2a). dI/dV spectra for analogous point defects in 2D-WS$_2$ (blue curve in Fig. 2b) show similar characteristics. The quasiparticle gap near the defect is the same as that measured on pristine 2D-WS$_2$ (black curve in Fig. 2b), and the defect dI/dV spectrum exhibits a characteristic defect state below the valence band, similar to the defect in MoSe$_2$. The inset in Fig. 2b shows a spatially resolved conductance scan, dI/dV (x,V), along the line crossing the point defect in the panel. The line scan reveals the spatial extent of the described spectroscopic features within 2 nm of the center of the defect. The gap edges are determined by taking the logarithm of the dI/dV, as described in ref.24[24]. The absence of any IGS resonance associated with the defect is in stark contrast to previous expectations for chalcogen vacancies[4–7,10,17] in both MoSe$_2$ and WS$_2$.

To validate our observation, we perform a series of control experiments on both MoSe$_2$ and WS$_2$ to exclude various scenarios that might prevent IGS from being observed by STS. We are able to rule out orthogonality of the wavefunctions of the tip and defect states leading to a vanishing tunneling matrix element; dynamic and static charging of the defect and the influence of the graphene substrate (see Supplementary Information for details). The control experiments on both MoSe$_2$ and WS$_2$ establish that IGS are not associated with the most abundant type of defects in our TMD samples.

**Theoretical approach and comparison with the experiments.**

Since our STS data curves do not feature IGS, we turn to first-principles DFT and GW calculations for further insight. Prior calculations of S vacancies and substitutional point defects[4–7,10,17,18,28,29,34–36] in 2D-MoS$_2$ have shown that defect energy levels can vary significantly, depending on whether the defect is a vacancy or a substituted atom. Based on growth conditions for both single layers (MoSe$_2$ and WS$_2$) and the sample treatment prior to our scanning probe measurements, oxygen, carbon, silicon, nitrogen, and hydrogen may potentially substitute chalcogen atoms. We can exclude S and Se substituents in, respectively, MoSe$_2$ and WS$_2$ samples, as they are grown in different experimental set-ups, avoiding cross contamination of chalcogen sources. Additionally, since S and Se atoms possess similar van der Waals radii they would result in similar nc-AFM images and would not be expected to appear as a depression in chalcogen lattice. According to prior DFT calculations, hydrogen, nitrogen, carbon and silicon substituents form IGS within the semiconducting gap[10,18] (See also Supplementary Information, Figure S5). Oxygen substitution, on the other hand, is predicted to suppress the deep in-gap states associated with the vacancies[18,28], and oxygen is present during the CVD growth process of WS$_2$. While the MBE samples are Se-capped to prevent air exposure during the transfer to the STM set-up, desorption of atmospheric H$_2$O, O$_2$, or CO$_2$ from the air-exposed sample holder is likely to occur during the annealing process for decapping (See Methods for further description). Recent DFT calculations propose O$_2$ molecules to chemisorb and dissociate on the chalcogen vacancies[29,37], leading to a stable system formed by a O substitutional chalcogen and a O adatom, further supporting the likelihood that O may be incorporated into the MBE samples. Pristine sulfur vacancies can, however, be generated during annealing in vacuum[10,38], as they present low formation energies[17,18]. By annealing our WS$_2$ samples at about 600 C in ultra high vacuum (UHV) conditions, we observe the formation of sulfur vacancies. The STS spectra of the V$_S$ in WS$_2$ reveals a characteristic fingerprint with two narrow unoccupied defect states[39]. The identification of in-vacuum generated V$_S$, with a robustly different electronic structure than O$_S$, validates the assignment of the observed defects in Fig. 1 as substitutional oxygen.

Standard DFT calculations are well known to underestimate bandgaps and quasiparticle energy levels[40]. For point defects in particular, DFT can incorrectly predict the relative energies of localized defect and delocalized bulk states[41,42]. Our TMDs feature both defect states, which are localized near and at the defect, and non-defect extended states associated with the pristine system[5,6,17]. To compute the energies of both defect and extended TMD states with spectroscopic accuracy, we use *ab initio* many-body perturbation theory within the GW approximation[43,44], correcting standard DFT energies with additional many-electron self-energy effects relevant to charged excitations in these systems. Our GW calculations are expected to accurately predict the energies of both localized and extended states, and has previously been shown to predict accurate bandgaps and electronic structures of pristine TMDs[45–47].

We first relax the atomic coordinates of a monolayer of MoSe$_2$ with several different point defect types, including a Se vacancy and substitutional atoms, using DFT within the local density approximation (LDA)[48] (see Methods). We then use a force-field model to simulate nc-AFM images, following a previously established method by Hapala, *et al*.[32] Figures 3 a-c show the relaxed atomic structure of a Se vacancy (**V**$_{Se}$, Fig. 3a), and two different substitutional (oxygen, **O**$_{Se}$, -Fig. 3b- and hydrogen, **H**$_{Se}$ – Fig. 3c) Se defects in MoSe$_2$. The corresponding simulated nc-AFM images from these three types of point defects are shown in Figs. 3d-f when the defect is located in the top layer (facing the tip in our experimental system) and in Figs. 3g-i, for which the defect is placed in the bottom layer (facing the underlying graphene layer in experimental system). Interestingly, since both H and O substituents are recessed into the chalcogen layer, the simulated nc-AFM contrast in images from all three types of atomic defects is equally compatible with the measurements in Fig. 1a. The atomic sized depression is assignable to the missing Se atom or the substituent atom in the upper Se-sublattice facing the tip. An apparently protruding Se atom can be identified as the equivalent defect in the bottom Se-sublattice facing the underlying graphene layer; the bottom defect slightly pushes the Se atom upward relative to the pristine lattice. In 2D-WS$_2$ we identified the counterpart defects on the top and bottom S-sublattices, which exhibit the same nc-AFM morphology as described for MoSe$_2$ (see Supplementary Information, Figure S1).

To identify which defects are likely observed in Fig. 1, we turn to calculations of their quasiparticle electronic structure using *ab initio* many-body perturbation within the GW approximation[43,44] since STS measures the energies of quasiparticle excitations, i.e., addition or removal of an electron. We perform DFT and GW calculations for both the bare Se vacancy and aforementioned substituted Se defects in MoSe$_2$. (Prior calculations of the bare chalcogen vacancy in WS$_2$ suggest that the defect electronic structure is qualitatively similar in both TMDs[49,50].) We emphasize that *ab initio* GW calculations are crucial to accurately describe the energy levels of both defect and non-defect states, required for comparing our computational results with the experimental observations and defect identification. As shown in Fig. 4a, the bulk GW gap between the valence and conduction band edge states is 2.2 eV. Our GW calculations of the Se vacancy reveal a deep, doubly-degenerate IGS located near the experimental Fermi level (red line), in qualitative agreement with prior DFT results. (Our DFT calculations of the H substitution Se defect also show IGS states, see Supplementary Information, Figure S5.) On the other hand, our *ab initio* GW calculations of the substitutional O defect show no IGS, which can be rationalized by the fact that O is isoelectronic to Se and S, and the resulting band gap of 2.1 eV compares well with the experimental gap. The lack of IGS for the substitutional O defect, together with its simulated nc-AFM image, is consistent with the point defect shown in Fig. 1.

We compare the measured spatial distributions of the LDOS around the top O$_{Se}$ defect and the calculated wavefunctions for both the Se vacancy and the O$_{Se}$ defect. Figure 4b shows a representative dI/dV constant-height conductance map measured at the CB (V$_{sample}$= 0.7 V) of the top O$_{Se}$ defect for MoSe$_2$. The calculated LDOS spatial distribution near the CB edge for both the Se vacancy (Fig. 4c) and O$_{Se}$ defect (Fig. 4d) closely resembles the experimental map. Similarly, the simulated LDOS of the vacancy (Fig. 4f) and top O$_{Se}$ defect (Fig. 4g) at the VB edge agree well with experiment, as shown in Fig. 4e. This result reveals limitations of STM imaging for discerning between the two types of defects, as previously addressed in the literature[13,20,21,26,27]. We compare the measured STM image at the CB edge for MoSe$_2$ with the calculation of the vacancy at both the energy of the deep in-gap state and the CB edge. The simulated STM image for the vacancy at the IGS (see Supplementary Information, Figure S4) is distinct from the experimental dI/dV maps in shape and registry to the MoSe$_2$ lattice. We emphasize that access to the position of the Mo sublattice from nc-AFM imaging enables a direct comparison with the calculated wave functions of the defects. This information is crucial due to the similar symmetries of the V$_{Se}$ IGS wavefunction and the CB wavefunction associated with from V$_{Se}$ and O$_{Se}$ defects.

**Discussion**

Taken together, our analysis suggests that the most commonly observed point defects in MoSe$_2$ and WS$_2$ are O substitutions at Se and S sites, respectively. A number of TEM and STM studies have investigated point defects on various TMD samples[9,17], and the majority of chalcogen-site point defects have been identified as vacancies. In addition, inferring the contribution of a specific type of defect to the general response of the material based on transport or photoresponse measurements might be challenging, as the averaged impact from other defects cannot easily be excluded. Although oxidation has been observed in TMD semiconductors[16] and the absence of IGS due to oxygen substitution has been discussed based on DFT calculations[29], no direct experimental access to the electronic structure of the individual oxygen-related defect has been reported so far. As mentioned, the interpretation of atomically resolved STM images of 2D-TMDs is not straightforward due to the convolution of structural and electronic effects[24], leading to a non-consistent interpretation of the defect type across the current literature[13,16,20,21,26,27]. In the case of TEM, the low threshold for electron beam-induced damage in TMDs[22] make it difficult to identify intrinsic vacancies from new vacancies created by electron beam irradiation[3]. Furthermore, light elements such as C and O contribute only weakly to TEM image contrast. While there is a clear difference between an extant chalcogen and a top or bottom vacancy, the difference between a chalcogen vacancy and an oxygen substitution is quite subtle[51] and could only be resolved in high signal to noise images, which require correspondingly high radiation doses that introduce high numbers of new defects. Furthermore, the lack of direct access to the electronic structure of individual defects by TEM hinders further and direct differentiation between defects presenting similar contrast. We further note that the annealing treatment used here extends not only to MBE or CVD grown samples for their characterization in UHV conditions, but also to transferred samples in order to remove contamination caused by air exposure or residues from the transfer process[16,17,35]. Therefore, our conclusions about the prevalence of substitutional oxygen in these 2D-TMDs are expected to be quite general. Further, as the presence of IGS has been connected to key photophysical properties of these materials, the identification of the nature of these common defects will advance efforts to control functionality in the important 2D-TMD class of systems.

In conclusion, we use a combination of experiment and theory to identify substitutional oxygen as a prolific point defect in 2D-MoSe$_2$ and 2D-WSe$_2$, directly correlating atomic structure and local spectroscopy. Relevantly, we show how neither of the described isolated methods - nc-AFM, STM and STS - could reveal the structure of the most abundant type of defect in our 2D-MoSe$_2$ and 2D-WS$_2$ samples. Our calculations predict that oxygen substituted in the chalcogen sublattice does not form deep in-gap states, consistent with STS measurements. Our findings suggest that substitutional oxygen point defects, and not just chalcogen vacancies, will play an important role in determining TMD photophysics and guide current efforts of the scientific community towards device functionality.

## Methods

**Experimental details**

Single layers of MoSe$_2$ were grown by molecular beam epitaxy on epitaxial bilayer graphene (BLG) on 6H-SiC(0001). The growth process was the same as described in Ref.52[52].The structural quality and the coverage of the MoSe$_2$ samples were characterized by in situ reflection high-energy electron diffraction (RHEED), low-energy electron diffraction (LEED), and photoemission spectroscopy (PES) at the HERS end station of Beamline 10.0.1, Advance Light Source, Lawrence Berkeley National Laboratory. The WS$_2$ films were grown on epitaxial graphene/SiC substrates by a modified chemical vapor deposition process at T = 900°C which uses H$_2$S as chalcogen source and WO2.9 powder as metal source as described in detail in ref. 53[53]. The data discussed in the manuscript has been reproducible measured over different sets of MoSe$_2$ and WS$_2$ samples, grown with the described methodology. The samples investigated here were prepared under fundamentally different conditions. Whereas the MoSe$_2$ samples were grown by MBE from elemental sources and experienced UHV conditions in which they were subsequently capped by a protective thin layer of Se, the CVD WS$_2$ samples were grown from a metal oxide precursor and H$_2$S gas and exposed to air. There are several stages in which the oxygen substitution could be introduced: during the growth itself (particularly for the CVD sample), under ambient conditions (atmospheric H$_2$O, O$_2$, or CO$_2$ could be potential reactants) or while annealing in vacuum previously adsorbed molecules on vacancy sites that could split and leave the O behind.

STM/nc-AFM imaging and STS measurements were performed at T = 4.5K in a commercial Createc - UHV system equipped with an STM/qPlus sensor. STS differential conductance (dI/dV) point spectra and spatial maps were measured in constant-height mode using standard lock-in techniques (f=775Hz, V$_{r.m.s.}$=2.1mV, T =4.5K). dI/dV spectra from Au(111) were used as an STS reference to control tip quality. Nc-AFM images were recorded by measuring the frequency shift of the qPlus resonator (sensor frequency f$_0$=30kHz, Q= 25000) in constant-height mode with an oscillation amplitude of 180 pm. Nc-AFM images were measured at a sample bias V$_s$=-50mV, using a tip functionalized with a single CO molecule. STM/STS data were analyzed and rendered using WSxM software[54].

**Theoretical details**

Calculations proceed in two steps. First, we perform DFT-LDA calculations for a single vacancy or substitutional defect in a large 5x5 supercell, corresponding to a small defect concentration of 2%. In this large supercell, which includes 74 atoms and 15 Å of vacuum, the defect density is low enough so that interactions between them can be safely neglected. Our DFT-LDA calculations use norm-conserving pseudopotentials, and we explicitly treat of 4s and 4p semi-core electrons in Mo. (Full details of our DFT-LDA calculations appear in the SI.) Second, we perform one-shot $G_0W_0$ calculations on our relaxed defect structures, starting from DFT-LDA. To increase accuracy, our GW calculations rely on a fine non-uniform sampling of reciprocal space[55]. Since the defect system contains a mixture of localized and extended states, we explore different treatments of the frequency dependence of the dielectric response in the screened Coulomb interaction, W. We find that the qualitative picture remains the same regardless of whether the frequency dependence is treated in full[56–58] approximated by a Hybertsen-Louie generalized plasmon pole (HL-GPP) model,[43,44] or neglected entirely in the static limit. All GW results shown are obtained with the HL-GPP model[43,44]. Further computational details are provided in the SI.

**Acknowledgements**

This work was supported by the Center for Computational Study of Excited State Phenomena in Energy Materials (C2SEPEM), which is funded by the U.S. Department of Energy, Office of Science, Basic Energy Sciences, Materials Sciences and Engineering Division under Contract No. DE-AC02-05CH11231, as part of the Computational Materials Sciences Program. Work performed at the Molecular Foundry was also supported by the Office of Science, Office of Basic Energy Sciences, of the U.S. Department of Energy under the same contract number. S.B. acknowledges fellowship support by the European Union under FP7-PEOPLE-2012-IOF-327581. S.B. and M.M.U acknowledge Spanish MINECO (MAT2017-88377-C2-1-R). S.R.A acknowledges Rothschild and Fulbright fellowships. B.S. appreciates support from the Swiss National Science

Foundation under project number P2SKP2_171770. A.P. and O.V.Y. acknowledge support by the ERC Starting grant "TopoMat" (Grant No. 306504). M.F.C. acknowledges support from the U.S. National Science Foundation under project number EFMA-1542741. DFT calculations were performed at the Swiss National Supercomputing Centre (CSCS) under project s832 and the facilities of Scientific IT and Application Support Center of EPFL. This research used resources of the National Energy Research Scientific Computing Center (NERSC), a DOE Office of Science User Facility supported by the Office of Science of the U.S. Department of Energy under Contract No. DE-AC02-05CH11231 for the GW calculations. This research used resources of the Advanced Light Source, which is a DOE Office of Science User Facility under contract no. DE-AC02-05CH11231.

**Author contributions**

S.B., S.W., M.M.U., O.Y. and A.W.-B. conceived the initial work. S.B., S.W. and A.W.-B. designed the experimental research strategy. S.B., S.W. and B.S. performed the STM/STS/nc-AFM measurements. A.W.-B. supervised the STM/STS/nc-AFM measurements. S.R.-A., D.Y.Q., S.G.L. and J.B.N. provided the DFT and GW theory support on defects on MoSe$_2$ and WSe$_2$. A.P. and O.Y. provided DFT calculation of defects on MoSe$_2$. H.R. and C.H. performed the MBE growth and characterization of the samples. S.-K.M. supervised the MBE growth. C.K., C.C., S.A. and A.S. grew the samples and developed the CVD vdW epitaxy process. M.M.U., M.F.C. and D.F.O. participated in the interpretation of the experimental data. S.B. wrote the manuscript with help from S.R.-A., B.S., D.Y.Q., D.F.O., J.B.N. and A.W.-B. All authors contributed to the scientific discussion and manuscript revisions.

**Data availability**

The data that support the findings of this study are available in the Supplementary Information and from the corresponding author upon reasonable request.

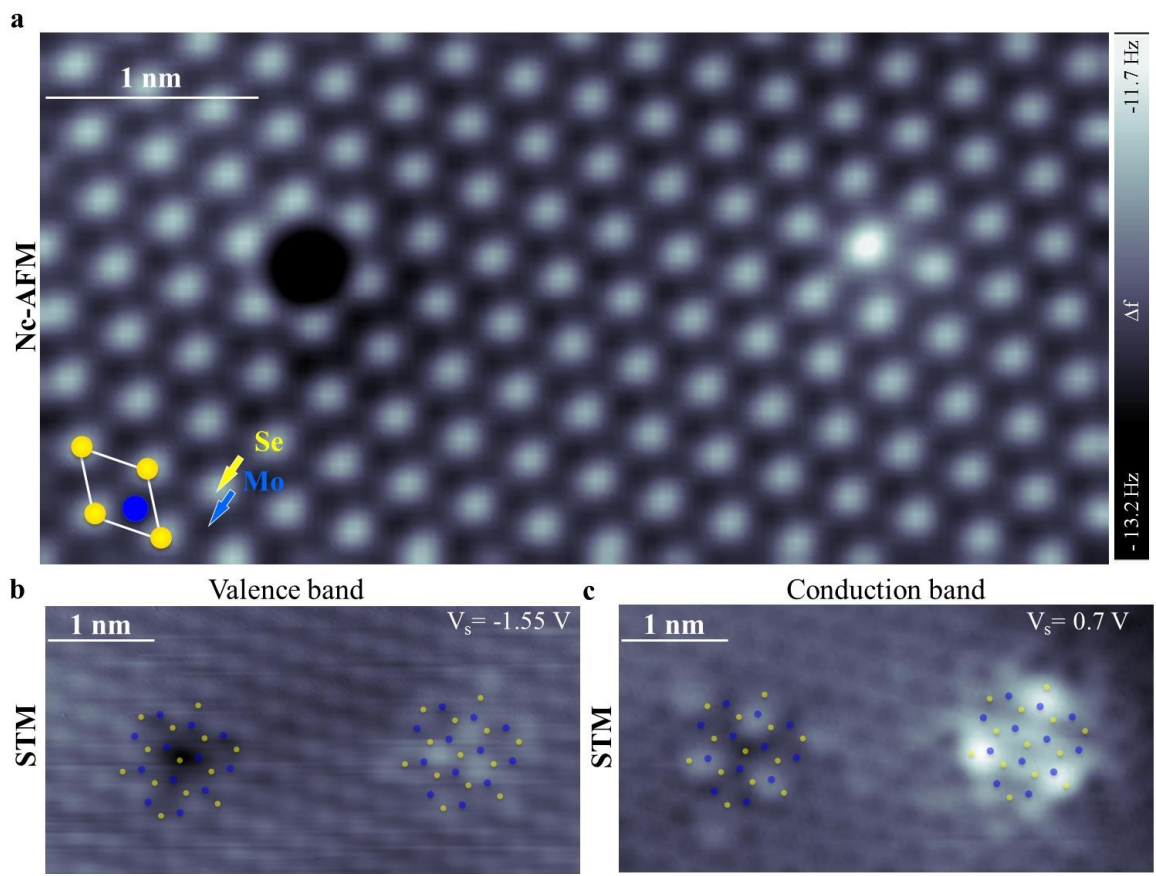

**Figure 1. Nc-AFM and STM images of the O$_{Se}$ top and bottom defects in 2D-MoSe$_2$. a,** CO-tip nc-AFM image of O$_{Se}$ top in the top Se layer of 2D-MoSe$_2$ (left) and O$_{Se}$ bottom in the lower Se layer facing the graphene substrate. Atomic resolution STM constant current images on the same area as in **a,** measured at the (**b**) valence and (**c**) conduction bands edges. Se (yellow dots) and Mo (blue dots) locations are indicated in the images.

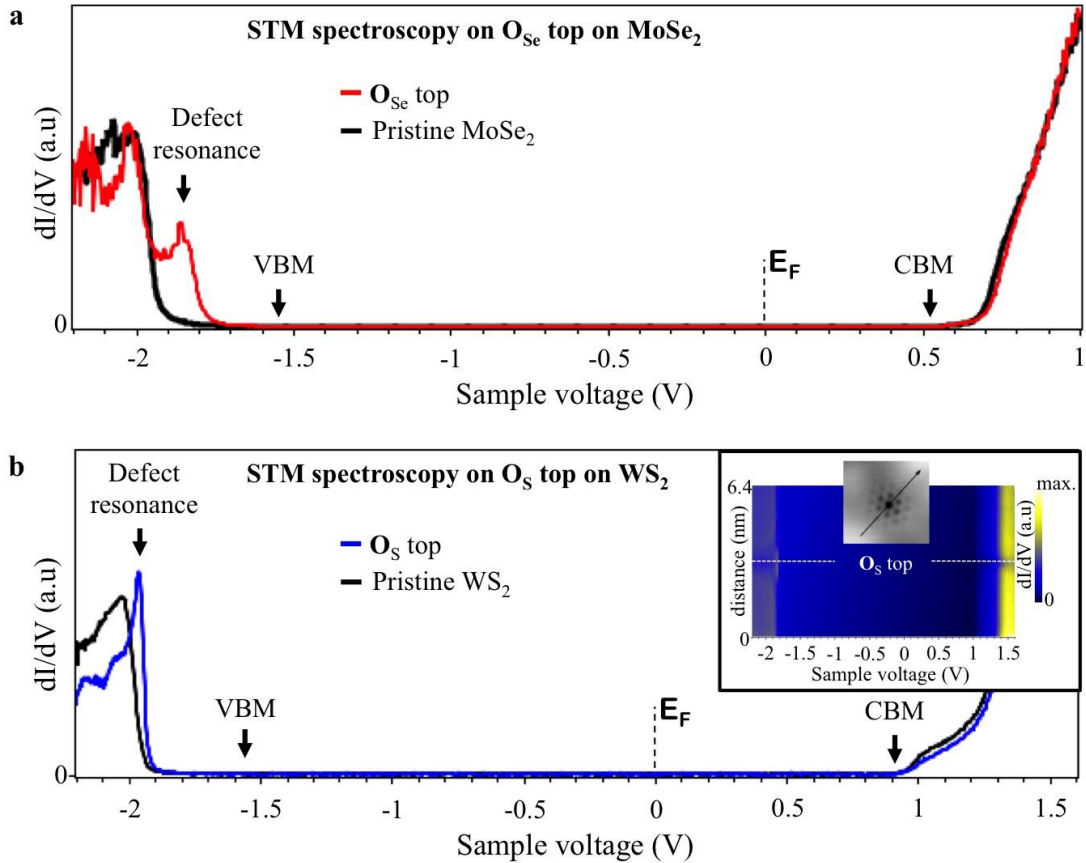

**Figure 2. Scanning tunnelling spectroscopy of substitutional oxygen in chalcogen site on 2D-MoSe$_2$ and 2D-WS$_2$. a,** Representative STM dI/dV spectra acquired on the left defect in Fig. 1 – substitutional oxygen at a Se site, **O$_{Se}$,** in 2D-MoSe$_2$ (red line) do not show deep in-gap states and a badgap equivalent to that measured on pristine sites (black line). Valence band maximum (VBM) and conduction band minimum (CBM) are marked with arrows. An additional defect resonance about 300 mV bellow the VBM it is observed in the defect's spectra. **b,** STM dI/dV spectra acquired on an substitutional oxygen, **O$_S$**, at a S site in 2D-WS$_2$ (blue line) also show an equivalent bandgap to that measured on pristine sites (black line), lack of deep in-gap states and a defect resonance deep inside the valence band. Inset: Spatially resolved dI/dV conductance scan across the **O$_S$** defect depicted in the inset reveals spatially distribution of the defects feature over 2 nm from the center of the **O$_S$** (dotted-white line). Sample voltage of 0 V represents the Fermi level (E$_F$).

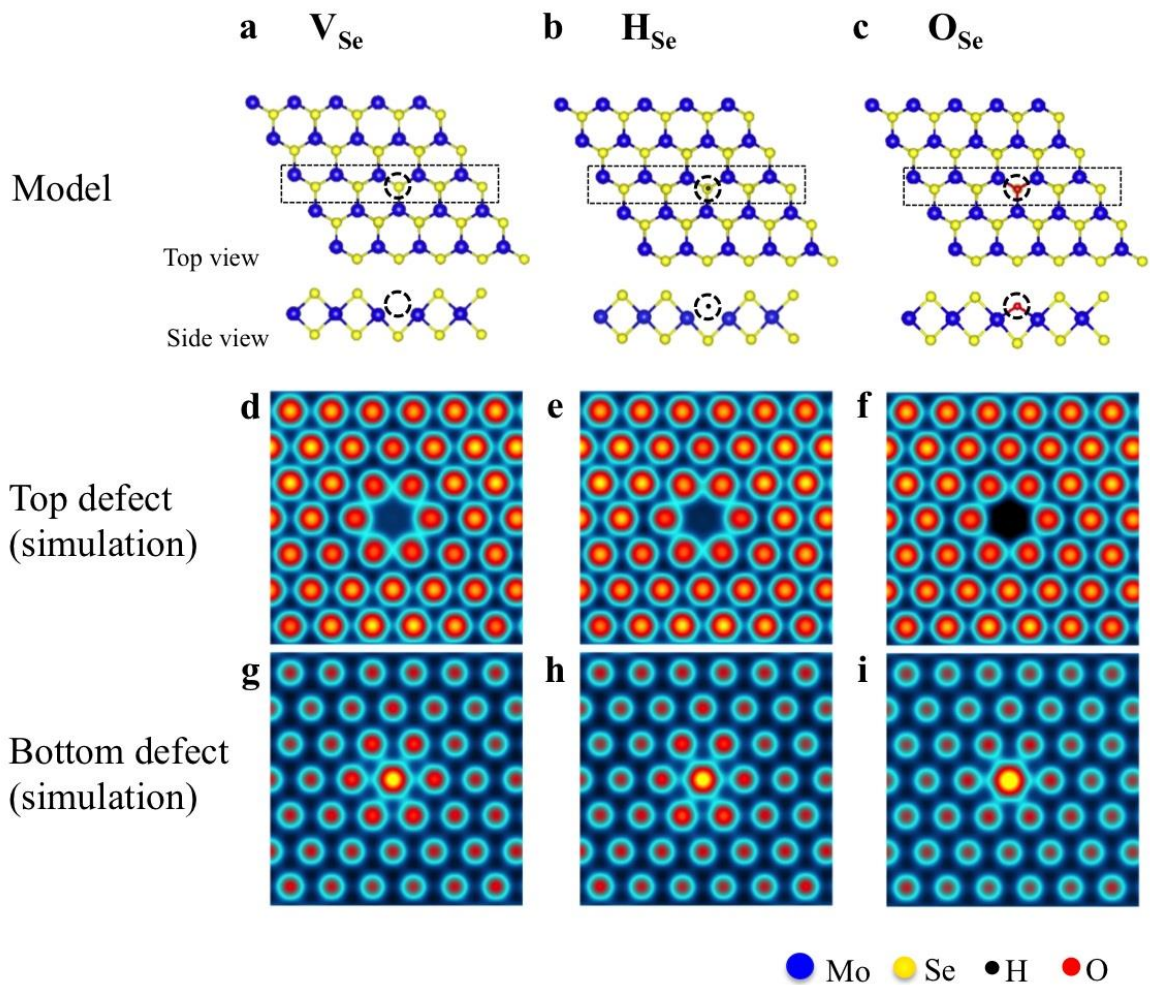

**Figure 3. Nc-AFM CO-tip simulation of a Se vacancy, H and O substitution.** Atomic structure from DFT-LDA relaxed coordinates of (**a**) a Se vacancy ($V_{Se}$), (**b**) hydrogen substitution ($H_{Se}$), and (**c**) oxygen substitution ($O_{Se}$) at a chalcogen site in a single layer of $MoSe_2$. Simulations of the nc-AFM images using a previous established method by Hapala *et al.*[32] of $V_{Se}$, $H_{Se}$, and $O_{Se}$ placed both (**d**-**f**) in the top layer (Se-sublattice facing the tip) and (**g-i**) in the bottom layer (Se-sublattice facing the underlying graphene layer), respectively.

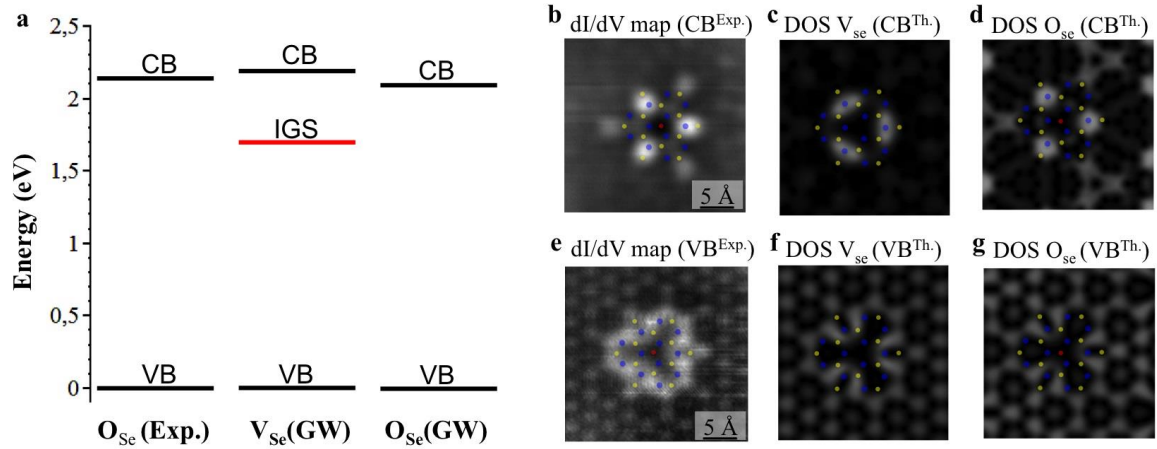

**Figure 4. Comparison of the band energy diagrams and the local density of states of pristine and O substitution Se defects in 2D-MoSe$_2$. a,** Band extrema energies extracted from the experimental dI/dV spectra in Fig. 2 (Exp.) are compared to the corresponding energies calculated using the GW approximation for the bare Se vacancy (**V$_{Se}$**) and a substitutional O at a Se site (**O$_{Se}$**). Energies of valence (VB) and conduction (CB) bands are indicated by black lines; while the deep in-gap states (IGS) appearing in the V$_{Se}$ is indicated by the red line, which dictates the Fermi level of the calculated system. To aid comparison, all VB energies have been set to zero. The three-fold symmetry and spatial extent observed in the experimental dI/dV constant-height conductance map measured at the (**b**) CB energy (V$_{sample}$=0.7 V) are seen on both (**c**) the pristine V$_{Se}$ and (**d**) the **O$_{Se}$**. Similarly, the experimental spatial extent at the (**e**) VB (V$_{sample}$=-1.5V) also reproduced the simulated LDOS of both (**f**) V$_{Se}$ and (**g**) **O$_{Se}$**.

# Supplementary Information

**Identifying substitutional oxygen as a prolific point defect in monolayer transition metal dichalcogenides with experiment and theory**

Figure S1 shows a typical STM image on 2D-WS$_2$ growth by chemical vapor deposition (CVD) on multilayer graphene (MLG) on SiC. We observed two different types of individual point defects with six-fold symmetry: a daisy-like shape (left defect in Fig. S1(b) and Fig. S1(c)); and donut-like shape (right defect in Fig. S1(b) and Fig. S1(d)). Supported by nc-AFM images acquired with a CO-functionalized tip we identify two O substituted sulfur atom (O$_S$): (1) The one atom size depression (left defect in Fig. S1(e) and Fig. S1(f)) indicates an apparently missing S atom in the upper S-sublattice facing the tip; and (2) the apparently increased height of a S atom (right defect in Fig. S1(e) and Fig. S1(g)) can be attributed to a protruding S atom in the top S layer in response to a O$_S$ defect in the bottom S-sublattice facing the underlying graphene layer. The described relaxation agrees with DFT geometry relaxation of the defect structure, similar to the O$_{Se}$ defect in MoSe$_2$ shown in Fig. 3(a)-(c).

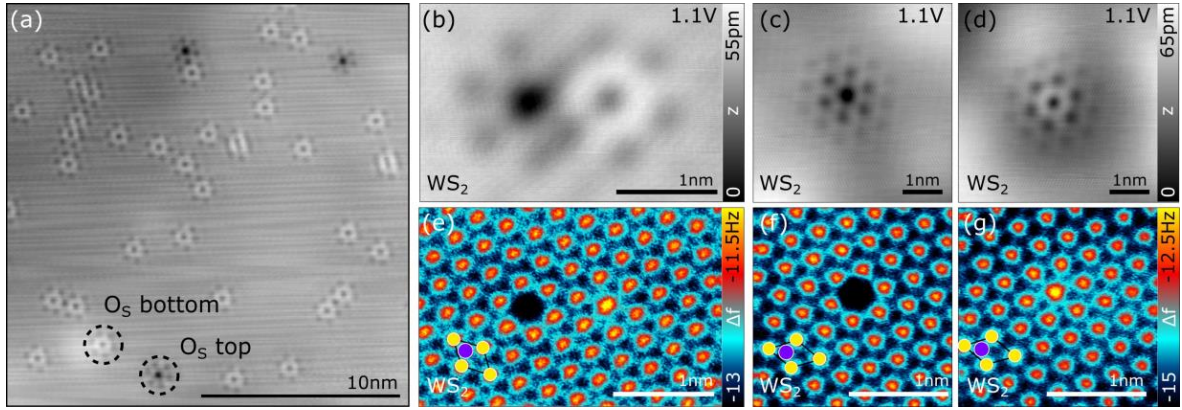

**FIG. S1**. (a) Large scale STM image of a WS$_2$ region showing a high density of point defects. The two different types of defect symmetry can be identified as single O-substituted S atoms placed either in the top (O$_S$ top) or in the bottom (O$_S$ bottom) S-sublattices. (b) Close-up view STM image at the conduction band of an O$_S$ top (left defect) and O$_S$ bottom (right defect) next to each other. Notice that the STM contrast around each defect remains unaltered compared to the contrast observed for the isolated O$_S$ top (c) and O$_S$ bottom (d) defects. (e), (f) and (g) show the nc-AFM images measured on the same area and with the same tip as the STM images in (b), (c) and (d), respectively.

**Atomic-scale characterization of top and bottom substitutional oxygen in MoSe$_2$**

The 2D-MoSe$_2$ sample was grown by molecular beam epitaxy (MBE) on bilayer graphene (BLG) on SiC. In order to protect the MoSe$_2$ monolayer from contamination or adsorbates during the transport in air to the UHV-STM chamber, we deposited a Se capping layer with a thickness of about 10 nm, which subsequently was removed by annealing the sample at about 600K in the UHV-STM and then transfer to the STM for surface characterization.

We explore the most abundant type of point defects on MoSe$_2$ and identify substitutional oxygen in a Se atom site (O$_{Se}$) as argued in the paper and outlined in detail below. Figure 1 in the main manuscript shows the nc-AFM image of two different types of such O$_{Se}$ defects: (1) The one atom size depression in Fig. 1a on the left is a O$_{Se}$ in the upper Se-sublattice facing the tip (denoted O$_{Se}$ top); and (2) the apparently increased height of a Se atom on the right could be identified as a O$_{Se}$ in the bottom Se-sublattice facing the underlying graphene layer tip (denoted O$_{Se}$ bottom), slightly lifting the above laying Se atom facing the tip. We want to stress that the identification of the O$_{Se}$ top and bottom defect, and in general any type of defect in 2D-TMDs, is extremely challenging. We can only solve the defect structure by comparing nc-AFM measurements to nc-AFM calculations and scanning tunneling spectroscopy (STS) to advanced *ab initio* electronic structure calculations. Neither of those methods could reveal the structure assignment alone. Importantly, *ab initio* calculations were essential to establish a sound foundation for the structure assignment. Our structure assignment is based on the entirety of the following observations: (i) Both the top and bottom defect version exhibit an identical electronic defect signature as measured by STS (see Fig. S2 (a)). This is a strong indication that both defects belong to the same type of defect. (ii) The nc-AFM image reveals that the defect is located on a Se site with an apparently missing Se atom on the top surface layer in one case and a protruding Se atom in the other one (See Fig. 1). (iii) The simulated nc-AFM image of a Se vacancy and a subtituional defects considering H and O would, in principle, fit the observed nc-AFM contrast (see Fig. 3). This narrows down considerably the most probable defect possibilities. (iv) STS does not observe any evidence of deep in-gap defect states (see Fig. 2), which is expected for both a Se vacancy and the Substitutional H defect (see Fig. S5). Several options, including defect charging (see Fig. S3 and Fig. S5), or canceling of tunneling matrix elements due to symmetry reasons, that could prevent us from detecting such an in-gap state can be ruled out. (v) The spatial distribution of the calculated local density of states of $O_{Se}$ at the conduction and valence bad edge resemble the experimental dI/dV images at the corresponding voltages (see Fig. 4). (vi) We found an analogous behavior to all points discussed above for substitutional oxygen in $WS_2$ grown by chemical vapor deposition (Fig. 2, Fig. S1), which anticipates a general trend for semiconducting TMDs in general.

Note that similar STM features have been previously discussed as vacancies or substitutional atoms of different kinds in 2D-TMDs, indicating the challenging structure assignment of TMD defects and the current inconsistency in the community and the understanding of they real influence on their electronic properties of the materials[1–7].

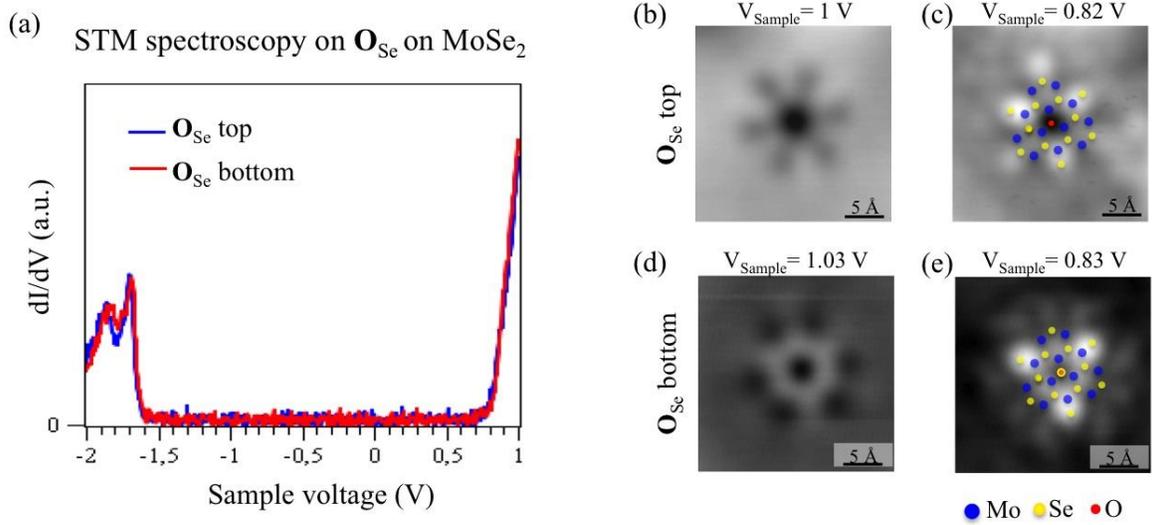

**FIG. S2.** (a) STM dI/dV spectrum acquired on an O substituent Se defect ($O_{Se}$) on MoSe$_2$, placed on the bottom Se-sublattice (red) shows the exact same spectroscopic features than the dI/dV spectrum acquire on the $O_{Se}$ placed on the top Se-sublattice (blue). The dI/dV spectra were measured with the same tip and in the same sample area. STM images from the (b) top $O_{Se}$ and (d) bottom $O_{Se}$ at voltages about 200mV above the conduction band, respectively, show a six-fold symmetry, instead of the three-fold symmetry characteristic from the STM images on the (c) top $O_{Se}$ and (e) bottom $O_{Se}$ at the conduction band edge.

**Challenging the presence of an in-gap defect state**

As outlined in the manuscript, there are several known mechanisms that can preclude detecting a localized in-gap state to be detected by STS, particularly due to dynamic or static charging of the defect. In Figure S3 (a) we compare STS spectra taken at a typical tunneling set point of I = 3nA at V = 1V with spectra taken at about 3Å closer. Apparently, there are no additional defect resonances observed but additional peaks associated with direct tunneling into the underlying graphene. This suggests that tunneling from a deep in-gap state to the graphene should be readily established since even a tip that is several Ångstroms farther away is able to directly tunnel into graphene with a considerable tunneling current of several nA at voltages as low as 10mV. This is true unless the charged defect is stabilized by atomic relaxations in the film. Atomic relaxations on ionic films have been shown to stabilize different charge states of individual atoms[8]. Different charge states can be discriminated[9] and charge state switching[10] detected by Kelvin probe spectroscopy. Fig. S3 (b) shows the Kelvin probe parabolas measured on the $O_{Se}$ (blue) and on the bare substrate (green). In the bias range probed no charge state switching was observed. In addition, the local contact potential difference corresponding to the bias at the vertex point indicates that the defect is neutral.

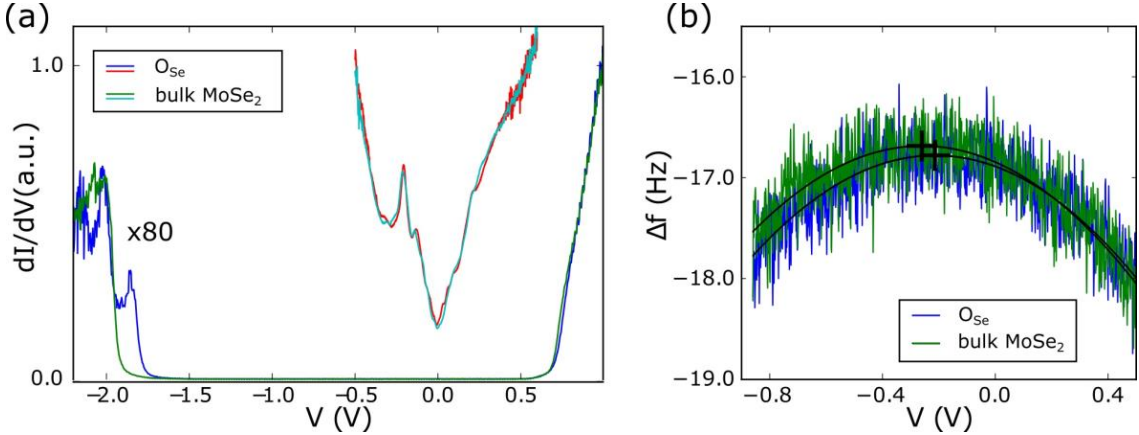

**FIG. S3.** (a) dI/dV spectra recorded on the $O_{Se}$ top (blue, red) and on the bare substrate (green, cyan). The spectra were taken at different tip heights: at a set point of I = 3nA at V = 1V (blue, green) and I = 600nA at V = 0.6V (red, cyan), respectively. (b) Kelvin probe parabolas recorded on the $O_{Se}$ top (blue) and on the bare substrate (green).

To further test this scenario, we compare the energy-dependent spatial distribution of the DOS around the defect we observe in the experiment with the simulated ones for the case of both, the bare Se vacancy and the O substitutional defect. Figure 4 shows the experimental dI/dV constant-height conductance maps, compared to simulated images using DFT wavefunction slices of $V_{Se}$ and $O_{Se}$ valence and conduction band edges. The charge distributions show a three-fold symmetry, extending spatially over two (pristine) lattice constants from the defect. In Fig. S4 (c) we show the spatial distribution of the DOS from the theoretically predicted in-gap state for the pristine $V_{Se}$, which exhibits a similar symmetry, but does not fit the experimentally observe registry to both Se- and Mo-sublattice at the defect conduction band, excluding the possibility of an energy shift due to charging of the in-gap state into the defect conduction band.

Finally, the fact that both the top and bottom chalcogen vacancies are electronically equivalent (Fig. S2 (a)) represents a strong indicative of the negligible influence of the graphene substrate and defect-substrate interactions.

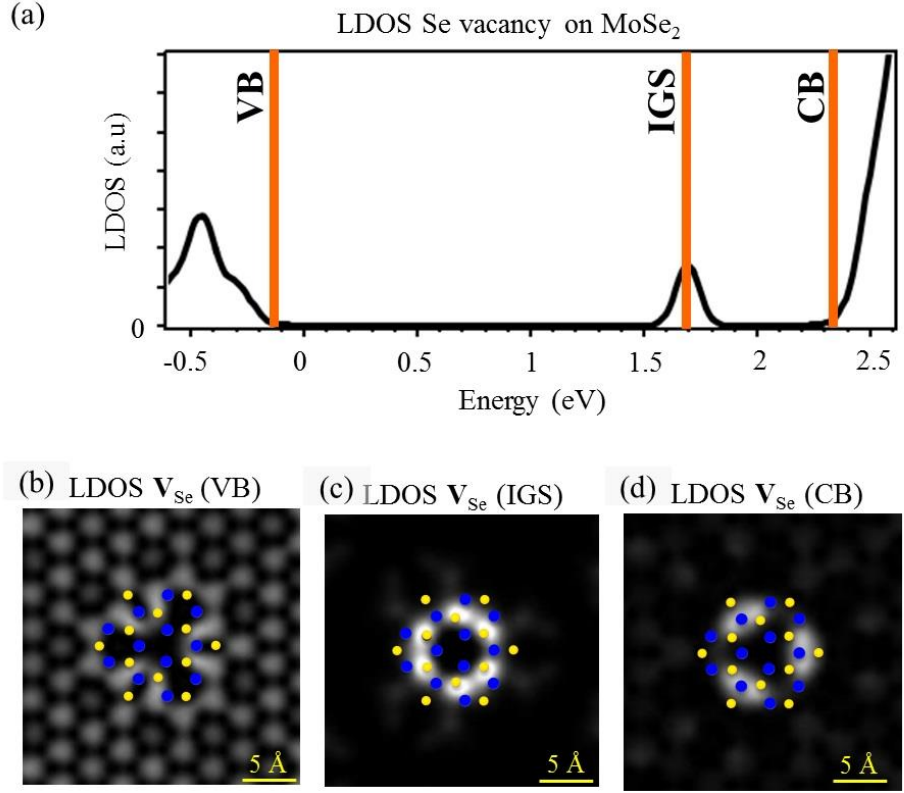

**FIG. S4.** (a) Calculated LDOS of monolayer MoSe$_2$ with Se vacancies within the GW approach. The Fermi level (0 eV) is placed at the valence band onset. We observe a gap between non-defect states of 1.9 eV and the presence of a doubly-degenerated unoccupied in gap state, ~0.4 eV below the non-defect conduction band minimum. (b) Simulated LDOS from DFT wavefunction slice of the bare V$_{Se}$ valence band (VB), in-gap state (IGS) and conduction band (CB) edge. Calculated LDOS are compatible with the three-fold symmetry and spatial extent observed in the experimental images, at the defect site. However, the simulated LDOS of the predicted defect in-gap state does not resemble any shape we observed experimentally.

**DFT calculations on a various point-defects in monolayer MoSe$_2$ and their effect on the density of states**

To study the effect of point defects on the presence of localized in-gap state in monolayer MoSe$_2$, we perform Density functional theory (DFT) calculations. All DFT calculations were performed within the Quantum-ESPRESSO[11] package using the PBE exchange-correlation functional[12] and with norm-conserving pseudopotentials[11,13,14]. Plane wave basis set with a kinetic energy cutoff of 80 Ry was used for the wavefunctions. The atomic structures of supercells of up to 7x7 times the pristine unit-cell, containing the point defects were relaxed self-consistently with a 15x15 k-point mesh, and a monolayer-monolayer separation of at least 15 Å of vacuum in the out-of-plane direction. For the calculation of monolayer MoSe$_2$ on graphene, a slight tensile strain in the graphene layer (with a lattice constant of 2.5 Å) was imposed in order to match the 5x5 supercell of MoSe$_2$. The dispersion correction presented in Ref.15 [15] was included in this case in order to describe interaction between graphene and MoSe$_2$.

Figure S4 shows the resulting DFT density of states for the different configurations: a pristine monolayer as reference (black), a monolayer containing neutral Se vacancies (red), a monolayer containing hydrogenated Se vacancies, with hydrogen atom bound to the vacancy site (blue), a monolayer containing negatively charged Se vacancies (green) with an additional electron, and a monolayer containing neutral Se vacancy in a monolayer MoSe$_2$ deposited on graphene (orange). For all examined defects, DFT results suggest the appearance of defect-localized in-gap states, as discussed in the main text.

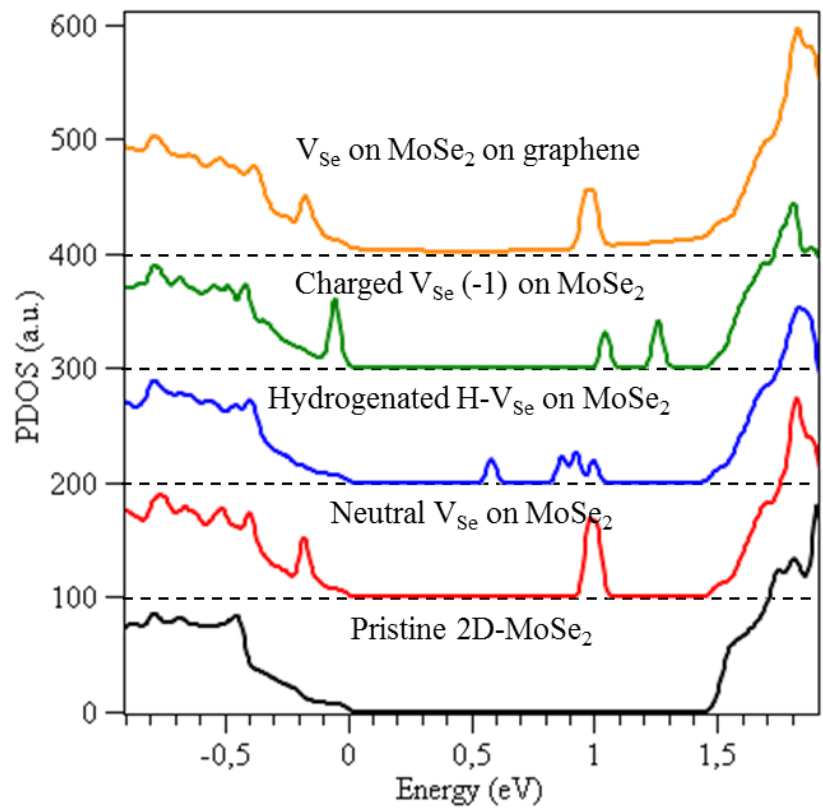

**FIG. S5.** Calculated DFT density of states (DOS) of pristine single-layer MoSe$_2$ (black line); and defective system containing one Se vacancy neutral (red line), hydrogenated (blue line), charged with one electron (green), and with the monolayer of MoSe$_2$ deposited on graphene (orange). All the spectra have been aligned with the conduction band edge at 0 eV.

**GW calculations of a Se vacancy and O substituent point defects in monolayer MoSe2**

We first performed DFT calculations within the local density approximation (LDA)[16] using the Quantum-ESPRESSO code[11]. The calculations were done on a 5x5 supercell arrangement of the MoSe$_2$ monolayer with either one Se vacancy or with a single O atom instead of one of the Se atoms. We used a plane-wave basis and norm-conserving pseudopotentials with a 70 Ry plane-wave cutoff. We included the Mo semi-core 4d, 4p and 4s states as valence states for our DFT and GW calculations. The distance between repeated supercells in the out-of-plane direction was 15 Å. We fully relaxed the geometry and included spin-orbit interactions as a perturbation. We performed the GW calculation within the BerkeleyGW code[17] using the generalized plasmon-pole model[18] and an energy cutoff of 25 Ry for the screening function. For the Se vacancy, we used 1677 unoccupied states, and tested the convergence with respect to unoccupied states by comparing to a calculation using 50,000 effective states obtained by averaging high-energy states within a small energy window. We verified convergence with respect to the number of bands, as discussed in ref. 19[19]. For the O substituent defect, we used 1245 and 1673 empty bands, and interpolated to the limit of infinite bands. For this defect we also verified the energy gap convergence at a higher screening cutoff of 35 Ry. In order to speed the convergence with respect to k-point sampling, we employed non-uniform sampling of the Brillouin zone, where the smallest q-vector corresponds to ∼1/1150th of a reciprocal lattice vector[20].